\documentclass[letterpaper]{article}

\usepackage{natbib,alifeconf}  
\usepackage{amsmath}
\usepackage{url,hyperref,cleveref}
\usepackage{booktabs}
\usepackage{amssymb}
\usepackage{subcaption}
\usepackage{float}
\usepackage{placeins}
\usepackage{balance}

\usepackage{xcolor}
\definecolor{dark-red}{rgb}{0.4,0.15,0.15}
\definecolor{dark-blue}{rgb}{0.15,0.15,0.8}
\definecolor{medium-blue}{rgb}{0,0,0.5}
\hypersetup{
    colorlinks, linkcolor={dark-red},
    citecolor={dark-blue}, urlcolor={medium-blue}
}

\newcommand\blfootnote[1]{%
\begingroup
\renewcommand\thefootnote{}\footnote{#1}%
\addtocounter{footnote}{-1}%
\endgroup
}

\usepackage{todonotes}
\title{Emergent Macro-Criticality from Micro-Critical Agents}

 \author{
     Nicolas Bessone$^{1,*}$ \and
     Erwan Plantec$^{1,2}$
     \\
     \mbox{}\\
     $^{1}$IT University of Copenhagen, Denmark \\
     $^2$ Barcelona Computational Foundation, Barcelona, Spain \\
     $^*$nbes@itu.dk
 } 

\begin{document}

\maketitle

\begin{abstract}

Criticality has been proposed as a key principle underlying complex behavior in biological and artificial systems; however, how criticality translates from individual dynamics to collective behavior remains unclear. We study this question using a multi-agent system with spatially constrained interactions in which agents sense neighboring light signals through exteroceptors and act by switching their own light on or off, thereby forming a dynamical interaction network at the macroscopic level. The agents' internal states are themselves governed by a reservoir dynamical system at the microscopic level. By varying the microscopic parameters around dynamical criticality, together with the macroscopic interaction topology, we systematically investigate the relation between the two levels. We find that near-critical dynamics within individual agents is not sufficient to produce collective critical-like avalanche statistics. Instead, scale-free behavior depends on the effective connectivity of the macroscopic interaction network, which controls activity propagation. As a result, macroscopic critical-like dynamics are enabled by microscopic regimes that deviate from criticality, with the required deviation depending on the properties of the interaction network. Investigating this relation, we find that slightly subcritical micro-level regimes support near-critical dynamics across a wider range of macroscopic parameters. These results show that in this multi-agent system, collective near-critical behavior depends on the interplay between internal dynamics and the interaction structure that governs activity propagation.
\end{abstract}


Code available at: \url{https://github.com/nhbess/emergent-macro-criticality}

\blfootnote{\textcopyright  2026 Nicolas Bessone and Erwan Plantec. Published under a Creative Commons Attribution 4.0 International (CC BY 4.0) license.}

\section{Introduction}

Criticality has been proposed as a unifying principle underlying complex behavior in biological and artificial systems. Evidence of near-critical dynamics has been reported across multiple domains, including neural activity, gene regulatory networks, and collective animal behavior, suggesting that operating near a critical point may support flexible and efficient information processing \citep{mora2011}. Criticality also holds important evolutionary implication as a selected property of organisms \citep{torres-sosa2012}.

Recent work has explored criticality in embodied agents and adaptive systems \citep{braccini2022,khajehabdollahi2022}, including reservoir-based approaches in which critical dynamics are embedded directly within the agent controller \citep{pontes-filho2025}, as well as studies showing that information processing in recurrent dynamical systems is maximized near the edge of chaos \citep{boedecker2012}. In parallel, studies of developmental and biological systems have shown that internal critical dynamics can influence higher-level organization \citep{kim_how_2018}. At a larger scale, criticality has also been investigated in collective systems, particularly in biological contexts, where scale-free correlations and collective responses have been observed \citep{romanczuk_phase_2022}.

Despite these advances, an important question remains open: how do critical dynamics propagate across scales? In particular, it is not yet clear whether near-critical internal dynamics at the level of individual agents are sufficient to generate critical behavior at the collective level, or whether additional constraints on interaction and connectivity are required. Developing a systematic understanding of emergent dynamical regimes and their relation to lower level properties is of particular importance. For example, such an understanding could shed important light on biological evolution and the selection pressures acting on lower level components such as cells in multicellular systems.

In this work, we address this question by studying a multi-agent system in which each agent contains an internal dynamical system capable of exhibiting critical dynamics, controlled by a single order parameter. Agents interact through spatially constrained sensing and signaling. This setting allows us to investigate how local dynamics and interaction topology jointly shape the emergence of collective behavior.

Our results show that near-critical dynamics at the level of individual agents are not sufficient to produce collective near-criticality. Instead, collective near-critical behavior depends on the ability of activity to propagate across the interaction network. We find that topological connectivity strongly shapes this propagation, and that the parameter region in which collective near-critical behavior is observed is displaced relative to the critical point of the agents’ internal dynamics. Moreover, slightly subcritical microscopic regimes support near-critical behavior across a wider range of macroscopic parameters.

\section[Experimental Methodology]{Experimental Methodology\footnote{For readers interested in an interactive presentation of the methodology, a companion webpage is available at \url{https://nhbess.github.io/nhbess/projects/live_viewer/about.html}}}

In order to study how critical dynamics propagate across scales, we distinguish between two levels of description in our system. At the \emph{micro level}, agents are endowed with an internal reservoir that governs how activity evolves within each agent. At the \emph{macro level}, agents interact through spatially constrained sensing, giving rise to collective dynamics at population level.

This separation allows us to explicitly control the dynamical regime at the micro level while observing its consequences at the macro level. In particular, we tune the internal reservoirs to be near their nominal critical branching point, a regime known to produce scale-free avalanche dynamics, and ask whether such near-critical dynamics persist, transform, or disappear when embedded in a spatially interacting multi-agent system.

\subsection{Micro Level: Reservoir}

The reservoir is a discrete-time stochastic recurrent network of $N_{\mathrm{neurons}}$ binary neurons designed to emulate the probabilistic activity propagation observed in neuronal avalanches. The formulation follows the branching-process interpretation commonly used to model avalanche dynamics \citep{beggs2003}.

At each timestep $t$ the state of neuron $i$ is $s_{t,i}~\in~\{0,1\}$, where $s_{t,i}=1$ indicates that the neuron is active. The network connectivity is represented by a transmission-probability matrix $W_{i,j} \in [0,1]$,  where $W_{i,j}$ is interpreted as the probability that neuron $i$, if active at time $t$, activates neuron $j$ at time $t+1$.

For every active neuron $i$, the model samples stochastic transmission events $X_{t,i,j} \sim \mathrm{Bernoulli}(W_{i,j})$. These variables represent whether neuron $i$ successfully propagates activity to neuron $j$ during that timestep. A neuron becomes active at the next timestep if it receives at least one successful transmission from an active pre-synaptic neuron. The update rule is therefore: 

\begin{equation}
    s_{t+1,j} =
    \left[
    \sum_{i=1}^{N_{\mathrm{neurons}}} s_{t,i} X_{t,i,j} \ge 1
    \right]
\end{equation}

Self-connections are typically removed, $W_{i,i}=0$, so that neurons cannot directly reactivate themselves in a single timestep. In this work we further assume a homogeneous transmission structure, meaning that all connections share the same propagation probability. The weight matrix therefore takes the form
\begin{equation}
W_{i,j} =
    \begin{cases}
    p^{\mathrm{micro}} & i \ne j \\
    0 & i = j
    \end{cases}
\end{equation}
where $p^{\mathrm{micro}} \in [0,1]$ is the common transmission probability governing activity propagation between neurons. 

Under the homogeneous connectivity assumption, each active neuron can potentially activate the remaining $N_{\mathrm{neurons}} - 1$ neurons, each with probability $p^{\mathrm{micro}}$. The expected number of activations generated by a single active neuron is therefore $(N_{\mathrm{neurons}} - 1) p^{\mathrm{micro}}$. The critical regime corresponds to the case in which, on average, one active neuron produces exactly one subsequent activation, yielding the critical value $p^{\mathrm{micro}}_{c} = \frac{1}{N{\mathrm{neurons}} - 1}$. 

\begin{figure}[t]
    \centering
    \includegraphics[width=1\linewidth]{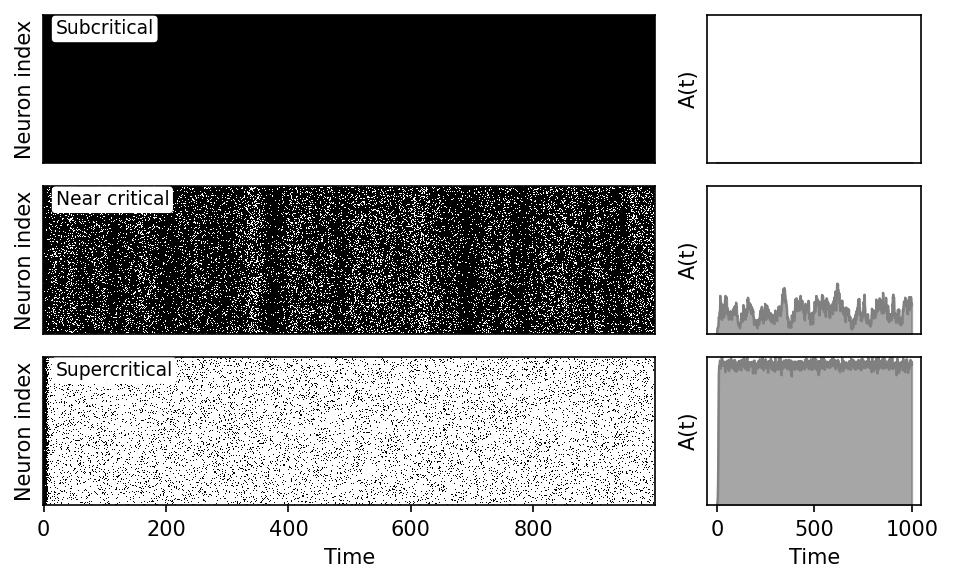}
\caption{Example of reservoir activity for different values of $p^{\mathrm{micro}}$, initialized from a single active neuron and evolved without external input. Left: neuron activity (raster plot). Right: population activity  $A(t) = \sum_i s_{t,i}$. Subcritical: activity quickly dies out. Near critical: self sustained activity. Supercritical: activity becomes dense and saturates.}
    \label{fig:reservoir}
\end{figure}

In this work, we use this branching point as a well-studied operating regime in which excitable networks exhibit balanced propagation and enhanced responsiveness to perturbations, including increased dynamic range and information-processing capacity \citep{kinouchi2006,shew2009,shew2013}. Fig.~\ref{fig:reservoir} illustrates the qualitative dynamical regimes of the reservoir as $p^{\mathrm{micro}}$ varies across subcritical ($p^{\mathrm{micro}}<p^{\mathrm{micro}}_{c}$), near-critical ($p^{\mathrm{micro}}\approx p^{\mathrm{micro}}_{c}$), and supercritical ($p^{\mathrm{micro}}>p^{\mathrm{micro}}_{c}$) regimes.

\subsection{Macro Level: Environment}

\begin{figure*}[t]
    \centering
    \includegraphics[width=1\textwidth]{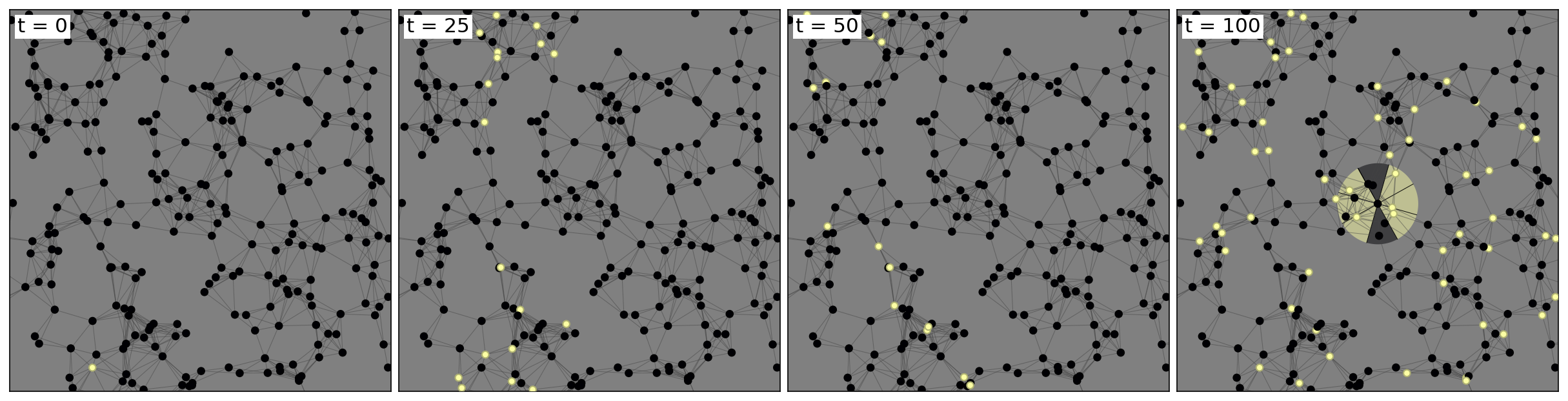}
    \caption{Spatiotemporal evolution of activity in an environment with $N_{\mathrm{agents}}=256$ agents, $p^{\mathrm{micro}} = 0.017962$, vision radius $v_r = 60$, and $E=8$ exteroceptor sensors. Starting from a localized perturbation, activity propagates through the interaction network, forming spatially distributed cascades. The final panel illustrates the sensory field of a representative agent.}
    \label{fig:environment}
\end{figure*}

The macro-level environment consists of a population of $N_{\mathrm{agents}}$ light-signaling agents embedded in a two-dimensional toroidal space. Agents occupy fixed spatial positions and interact exclusively through local visual perception of nearby neighbors, defining a static spatial interaction structure.

Each agent emits a binary light signal $l_{t,i} \in \{0,1\}$ indicating whether its light is OFF or ON at time $t$. At every timestep, each agent receives an egocentric sensory observation describing the angular distribution of illuminated neighbors within a finite vision radius $v_r$. This observation is discretized into $E$ exteroceptive sectors (see last panel at Fig.~\ref{fig:environment}), yielding a binary vector that indicates, for each sector, whether at least one visible neighbor with its light ON is present.

Agents do not have direct access to the internal states of others; all interactions are mediated through these local light-based observations. As a result, the global dynamics emerge from the coupling between spatial visibility and the agents’ signaling behavior.

Each agent is controlled by an internal reservoir network implementing the stochastic propagation dynamics described above. The $E$ sensory inputs are mapped onto $E$ designated input neurons of the reservoir, while a separate readout neuron determines the agent’s light state $l_{t,i}$. In this way, local observations perturb the internal reservoir dynamics, and the resulting activity drives the agent’s outward signaling.

\subsection{Simulation Protocol}

To investigate how micro-level reservoir dynamics influence macroscopic collective activity, we simulated populations of $N_{\mathrm{agents}}=256$ agents interacting in the environment described above. Agents were embedded in a two-dimensional toroidal domain of side length $L = 565$, ensuring periodic boundary conditions and eliminating edge effects, with positions drawn uniformly at random at the beginning of each run. Each agent contained a reservoir of $N_{\mathrm{neurons}}=64$ neurons, of which $E=8$ served as input nodes. 

Simulations were initialized with all agent lights in the OFF state except for a single randomly selected agent whose light was set to ON, thereby introducing a localized perturbation. Fig.~\ref{fig:environment} shows an example of a simulation run. The system was then evolved for $1000$ macro timesteps. At each macro timestep, every agent sampled binary light-based observations from neighbors within radius $v_r$, and these observations were clamped onto the corresponding reservoir input neurons. The reservoir dynamics then evolved for $5$ micro timesteps, after which the readout determined the agent’s next light state. In this way, $p^{\mathrm{micro}}$ controlled internal propagation within agents, whereas $v_r$ controlled the connectivity of the inter-agent interaction graph and thus the extent to which activity could spread collectively. 

To examine how the collective dynamical regime depended on internal propagation and spatial coupling, we performed parameter sweeps over the reservoir transmission probability $p^{\mathrm{micro}}$ and the vision radius $v_r$. For each parameter pair $(p^{\mathrm{micro}}, v_r)$, we run $200$ independent rollouts, each with a newly sampled spatial configuration of agent positions. Avalanche statistics were then extracted from the recorded binary light traces by partitioning each trajectory at time points where the population-wide light activity dropped to zero, such that each contiguous interval of nonzero activity was counted as one avalanche. Because the observable light activity can return to zero while latent activity persists in the internal reservoirs, a single rollout can contain multiple avalanches separated by periods of apparent silence. Consequently, the total number of extracted avalanches can exceed the number of rollouts and may vary across parameter settings even when the number of rollouts is held fixed.

\subsection{Avalanche Extraction}

Avalanches are identified from the macroscopic population activity signal. The activity at the collective at time $t$ is given by:

\begin{equation}
A_l(t)=\sum_{i=1}^{N_{\mathrm{agents}}} l_i(t)
\end{equation}

where $l_i(t) \in \{0,1\}$ denotes the light state of agent $i$ at time $t$. Formally, an avalanche begins at timestep $t_s$ when the system transitions from silence to activity, $A_l(t_{s}-1)=0$, $A_l(t_s)>0$, and ends at timestep $t_e$ when the system returns to silence, $A_l(t_e)>0$, $A_l({t_e+1})=0$. The avalanche duration is therefore $T=t_e - t_s + 1$.
The avalanche size measures the total amount of activity during the activity burst, and it is defined as

\begin{equation}
S=\sum_{t=t_s}^{t_e} A_l(t)
\end{equation}

Thus, duration captures how long the activity episode persists, while size measures the total number of agent activations occurring during the cascade. Avalanches that remain active at the final timestep of the recording are censored by the finite observation window, meaning that their true duration and size are unknown, and are therefore excluded from the analysis.

\subsubsection{Avalanche Observable Distributions}

To characterize the statistics of activity propagation, we analyze the distribution of avalanche sizes and durations across all runs.

Let $T_k$ and $S_k$ denote the duration and size of the $k$-th avalanche event, respectively, where $k \in \mathcal{K}$ indexes the set of all observed avalanches. The empirical probability distributions are estimated as

\begin{equation}
\begin{aligned}
P(T) &= \frac{1}{|\mathcal{K}|}\,\left| \{\, k \in \mathcal{K} \mid T_k = T \,\} \right| \\
P(S) &= \frac{1}{|\mathcal{K}|}\,\left| \{\, k \in \mathcal{K} \mid S_k = S \,\} \right|
\end{aligned}
\end{equation}

where $|\mathcal{K}|$ is the total number of avalanches observed. These distributions describe how frequently activity cascades of different temporal and spatial scales occur in the system. If activity typically dies out quickly, the distribution is concentrated at small durations. If the system frequently produces long avalanches, the distribution develops a heavy tail.

\subsubsection{Avalanche power-law fitting}

To assess whether the extracted avalanche observables exhibit scale-free statistics, we fitted discrete power-law models to the empirical distributions of avalanche duration $T$ and avalanche size $S$ following the statistical fitting framework of \citet{clauset2009}. Power-law statistics are a central prediction of critical branching processes and have been widely observed in neuronal avalanche dynamics \citep{beggs2003, mora2011}.

For a generic observable $x \in \{T, S\}$, representing avalanche duration or size, the fitted model assumes

\begin{equation}
P(x) = \frac{x^{-\alpha}}{\zeta(\alpha, x_{\min})},
\qquad x \ge x_{\min},
\end{equation}

where $\alpha$ is the scaling exponent, $x_{\min}$ is the lower cutoff of the fitted tail, and $\zeta(\alpha, x_{\min})$ is the Hurwitz zeta function, which normalizes the discrete power-law distribution over $x \ge x_{\min}$.

Empirical avalanche distributions rarely follow a power law over their entire range. Small avalanches are typically influenced by microscopic dynamics and finite-size effects, which can distort the scaling behavior. For this reason the model is fitted only to the tail of the distribution above a cutoff $x_{\min}$. The cutoff $x_{\min}$ is selected as the value that minimizes the Kolmogorov–Smirnov distance between the empirical cumulative distribution and the fitted power-law model. Once $x_{\min}$ is determined, the scaling exponent $\alpha$ is estimated by maximum likelihood over the tail $x \ge x_{\min}$. The goodness of fit is quantified using the Kolmogorov–Smirnov statistic

\begin{equation}
KS = \max_{x \ge x_{\min}} \left| F_{\mathrm{emp}}(x) - F_{\mathrm{model}}(x) \right|
\end{equation}

where $F_{\mathrm{emp}}$ and $F_{\mathrm{model}}$ denote the empirical and fitted cumulative distributions, respectively. Smaller values of $KS$ indicate better agreement between the observed data and the power-law hypothesis. The fitting procedure was implemented using the \texttt{powerlaw} Python package \citep{alstott2014}.


\subsection{Criticality Score}

To quantify the agreement between the observed avalanche statistics and the predictions of critical branching theory, we define a criticality score as a heuristic summary of agreement between avalanche shapes and branching-process expectations:

\begin{equation}
\mathcal{L}(p^{\mathrm{micro}},v_r) = |\alpha_S - 1.5| + |\alpha_T - 2| + KS_S + KS_T .
\label{equation:criticalityscore}
\end{equation}

Here, $\alpha_S$ and $\alpha_T$ denote the fitted power-law exponents for avalanche size and duration, while $KS_S$ and $KS_T$ are the corresponding Kolmogorov--Smirnov distances. The reference values $1.5$ and $2$ correspond to the classical critical exponents predicted by simple branching-process models, where the avalanche size distribution follows $P(S) \propto S^{-3/2}$ and the avalanche duration distribution follows $P(T) \propto T^{-2}$ \citep{beggs2003, mora2011, cramer2020}. Lower values of $\mathcal{L}$ therefore indicate avalanche statistics that are closer to the theoretical expectations of a critical branching process, while not by themselves establishing a macroscopic critical point in the statistical-mechanical sense.
\section{Results}

\subsection{Emergence of macro near-criticality from micro near-critical reservoirs}

\begin{figure}[t]
    \centering
    \includegraphics[width=0.95\linewidth]{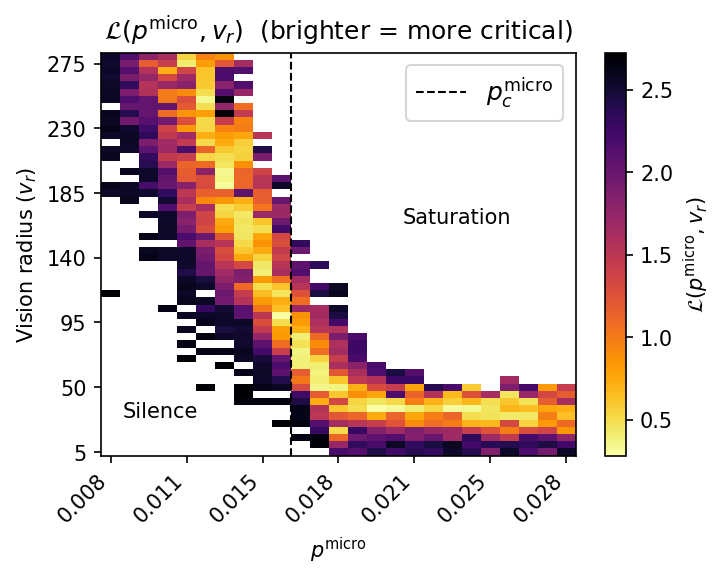}
    \caption{Heatmap of the criticality score $\mathcal{L}(p^{\mathrm{micro}}, v_r)$ across micro-level edge probability and vision radius. Brighter colors indicate more critical-like avalanche statistics. The dashed vertical line marks the theoretical critical point of the isolated reservoir $p^{\mathrm{micro}}_c$, shown for reference as it does not coincide with the macro-level critical regime. The system exhibits three regimes: silence, saturation, and an intermediate band where near-critical collective dynamics emerge.}
    \label{fig:heatmap}
\end{figure}

We first investigate whether bringing the internal reservoirs close to their branching critical point is sufficient, by itself, to generate non-trivial cascade dynamics at the collective level. Figure~\ref{fig:heatmap} presents the criticality landscape $\mathcal{L}(p^{\mathrm{micro}}, v_r)$ as a function of the micro-level edge probability and the agents’ vision radius.

The heatmap reveals a structured phase diagram at the collective level. For low $p^{\mathrm{micro}}$ and $v_r$, the system remains in a subcritical regime where activity fails to propagate and avalanches are either absent or too weak to yield reliable statistics, while for large $p^{\mathrm{micro}}$ and $v_r$, the system enters a saturated regime characterized by persistent activity and the disappearance of well-defined cascades.

Between these extremes, a narrow and curved band of high criticality scores emerges, corresponding to sustained cascade dynamics. This region is displaced from the theoretical micro-level critical point $p^{\mathrm{micro}}_c$ and shifts systematically with the vision radius, indicating that macro-level near-critical behavior arises only within a restricted region of parameter space shaped by spatial interactions governing activity propagation.

\subsection{Effective branching ratio $(\sigma_{\mathrm{eff}})$ and empirical shift}
The effective branching ratio provides a coarse measure of activity propagation across time, useful for identifying transitions between dynamical regimes. It defines a time-aggregated ratio of \textit{next-step activity} to \textit{current-step activity} over the total simulation duration $D$, computed as:

\begin{equation}
    \sigma_{\mathrm{eff}}
    =
    \frac{\sum_{t=1}^{D-1} A_l(t)}
    {\sum_{t=0}^{D-2} A_l(t)}
\end{equation}

This quantity should not be interpreted as a direct estimator of a branching ratio in the branching-process sense. It is computed from macro-level light activity, and its numerator and denominator differ only by boundary terms, namely the first and last activity values of the trajectory. Therefore, $\sigma_{\mathrm{eff}}$ is expected to approach $1$ whenever the total accumulated macro-level activity is large compared with these boundary contributions. We use $\sigma_{\mathrm{eff}}$ only as a coarse activity-flux balance measure, indicating where collective activity changes from dying out to being sustained.

\begin{figure}[t]
    \centering
    \includegraphics[width=0.95\linewidth]{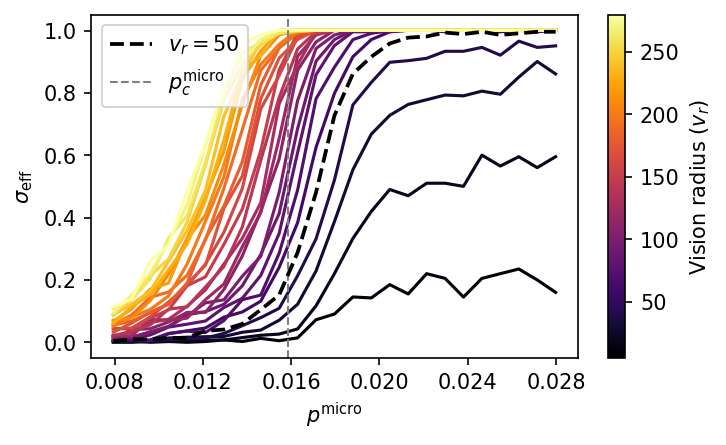}
    \caption{Effective branching ratio $\sigma_{\mathrm{eff}}$ as a function of the micro-level edge probability $p^{\mathrm{micro}}$ for different vision radii $v_r$ (color-coded). As $v_r$ increases, the transition of $\sigma_{\mathrm{eff}}$ toward $1$ shifts to lower values of $p^{\mathrm{micro}}$, indicating that spatial interactions enhance effective propagation. The dashed vertical line marks the theoretical critical point of reservoir $p_c^{\mathrm{micro}}$, while the dashed black curve highlights the transition between disconnected and connected interaction regimes occurs around $v_r \approx 47$ (shown for reference at $v_r=50$).}
    \label{fig:sigma_eff}
\end{figure}

Figure \ref{fig:sigma_eff} shows how $\sigma_{\mathrm{eff}}$ computed from macro-level activity varies with the micro-level propagation probability $p^{\mathrm{micro}}$ as a function of the vision radius $v_r$. As $v_r$ increases, the transition of $\sigma_{\mathrm{eff}}$ toward $1$ occurs at progressively smaller values of $p^{\mathrm{micro}}$, indicating that spatial coupling effectively enhances activity propagation at the macro level. This systematic shift suggests that the effective propagation of activity is not determined by internal dynamics alone, but is strongly modulated by the structure of inter-agent interactions.

\subsection{Shift of the macro-critical regime with vision radius}
\label{macrovsvr}

At the collective scale, propagation is constrained by the topology of the interaction network modulated by $v_r$. For a given vision radius, we define a graph where two agents $i$ and $j$ are connected if their distance is smaller than $v_r$. A key quantity is the size of the largest connected set of agents, known as the giant connected component. If $N_{\mathrm{GCC}}(v_r)$ denotes its size, we define

\begin{equation}
g(v_r)=\frac{N_{\mathrm{GCC}}(v_r)}{N_{\mathrm{agents}}}    
\end{equation}

which measures the fraction of the population that is mutually reachable through chains of interactions. When $g(v_r)$ is small, the system is fragmented into isolated clusters, and activity remains confined locally. As $v_r$ increases, the system undergoes a connectivity transition in which a large fraction of agents becomes connected, enabling activity to propagate across the population.

This transition occurs at a characteristic scale that can be predicted analytically. The interaction graph corresponds to a Random Geometric Graph, where $N_{\mathrm{agents}}$ are uniformly distributed over a domain of size $L \times L$ and edges are formed between agents within distance $v_r$. In this model, the expected number of neighbors is $k(v_r) = (N_{\mathrm{agents}}-1)\frac{\pi v_r^2}{L^2}$.

A classical result states that global connectivity emerges when the mean degree reaches $k \approx \ln N_{\mathrm{agents}}$. Solving for $v_r$ gives

\begin{equation}  
v_r^{\mathrm{conn}} = \sqrt{\frac{L^2 \ln N_{\mathrm{agents}}}{(N_{\mathrm{agents}}-1)\pi}}.  
\end{equation}

\begin{figure}[t]
    \centering
    \includegraphics[width=0.7\linewidth]{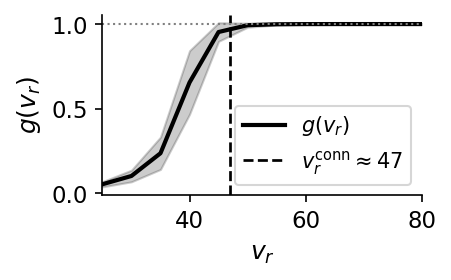}
   \caption{Topological transition in the interaction network as a function of the vision radius $v_r$. Giant component fraction $g(v_r)$, measuring the relative size of the largest connected cluster. A transition occurs around $v_r \approx 47$, corresponding to the emergence of global connectivity (dashed line). This transition marks the scale above which activity can propagate across a large fraction of the population rather than remaining confined to local clusters.}
\label{fig:connectivity}
\end{figure}

For our parameters ($N_{\mathrm{agents}}=256$, $L=565$), this yields $v_r^{\mathrm{conn}} \approx 47$, in agreement with the sharp transition observed in $g(v_r)$ in Fig.~\ref{fig:connectivity}. In this regime, the fragmentation of the interaction graph also explains why the effective branching ratio remains below unity for small $v_r$ (Fig.~\ref{fig:sigma_eff}). When the graph is disconnected, activity remains confined within local components and cannot propagate across the entire population.

Taken together, these results show that the shift of the macro-critical regime is governed by the interaction topology induced by $v_r$. Lower connectivity (small $v_r$) limits propagation, requiring higher values of $p^{\mathrm{micro}}$ to reach near-criticality. As connectivity increases, activity can spread more efficiently across the population, shifting the near-critical regime toward lower values of $p^{\mathrm{micro}}$. This systematic displacement reflects how the structure of interactions controls the effective propagation of activity at the macro scale. This provides a partial characterization of the micro–macro displacement in terms of topology: average degree controls the expected number of local interaction partners, while the giant connected component controls whether activity can propagate beyond local clusters.

\subsection{Avalanche statistics and deviations from mean-field scaling}
\begin{figure}[t]
    \centering
    \includegraphics[width=0.95\linewidth]{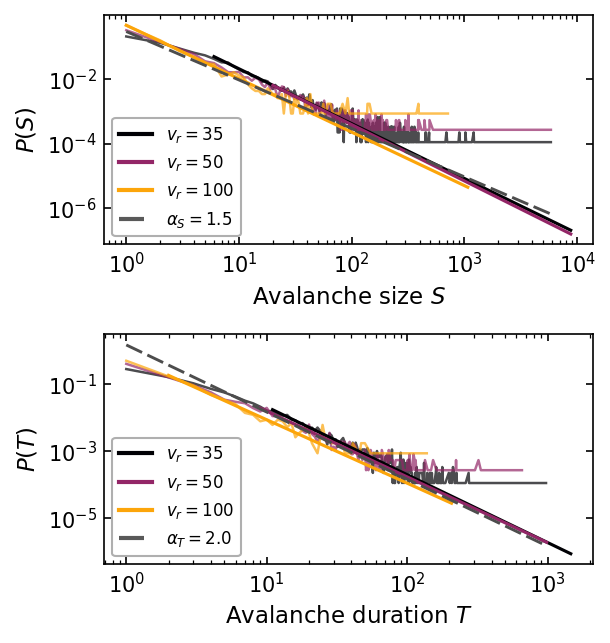}
\caption{
Avalanche size $S$ and duration $T$ distributions at the optimal connection probability $p^*$ for different vision radii $v_r$. Solid lines show the empirical distributions and the obtained power-law fits. The dashed reference lines correspond to the mean-field critical exponents $\alpha_S = 1.5$ and $\alpha_T = 2.0$ expected for critical branching processes.
}
    \label{fig:powerlaws}
\end{figure}

Figure~\ref{fig:powerlaws} shows the avalanche size and duration distributions at the most critical connection probability $p^*$ for different $v_r$. For each $v_r$, we identify $p^*$ as the value of $p^{\mathrm{micro}}$ that minimizes $\mathcal{L}$. The solid colored lines show the empirical probability $P(S)$ and $P(T)$ at $p^*$. The solid lines of matching color show the theoretical power-law fit. The dashed line indicates the mean-field reference slopes $\alpha_S = 1.5$ and $\alpha_T = 2.0$, predicted by branching process theory at criticality.

To assess whether the distributions are consistent with a power law, we apply the bootstrap goodness-of-fit test of \cite{clauset2009}. For each case, we generate $n=1000$ synthetic datasets from the fitted model, refit them, and compare their $KS$ statistics to the empirical one. The resulting goodness-of-fit value (denoted as $\mathrm{gof}_p$) corresponds to the fraction of synthetic datasets with larger $KS$ distance than the empirical data.

The results are reported in Table~\ref{tab:powerlaws}. The fitted size exponent $\alpha_S \approx 1.7$ consistently exceeds the mean-field prediction of $1.5$, despite the criticality score $\mathcal{L}$ being defined to favor that value. This indicates that, within our system, the configurations minimizing $\mathcal{L}$ tend to lie in a regime shifted away from the canonical mean-field branching expectation. Thus, while the observed avalanche distributions support scale-free-like or heavy-tailed cascade structure, they also indicate systematic deviation from the mean-field branching universality class. Similar exponent values have been reported in related reservoir-based and spatially structured systems \citep{pontes-filho2025}, suggesting that such deviations may reflect finite-size effects or constrained propagation rather than a peculiarity of the present setup.

\begin{table}[h]
\centering
\resizebox{\columnwidth}{!}{
\begin{tabular}{ccccccccc}
\hline
vd & $p^\star$ & $n$ & $\alpha_S$ & KS$_S$ & gof$_{p,S}$ & $\alpha_T$ & KS$_T$ & gof$_{p,T}$ \\
\hline
35  & 0.0196 & 8964 & 1.69 & 0.032 & 1.0 & 2.02 & 0.029 & 1.0 \\
50  & 0.0171 & 3715 & 1.74 & 0.035 & 0.9 & 1.96 & 0.033 & 1.0 \\
100 & 0.0155 & 1165 & 1.65 & 0.011 & 1.0 & 1.89 & 0.025 & 1.0 \\
\hline
\end{tabular}
}
\caption{
Power-law fit statistics for avalanche size $S$ and duration $T$ at the optimal connection probability $p^*$. Reported values include the fitted exponent $\alpha$, Kolmogorov--Smirnov distance $KS$, and goodness-of-fit measure $\mathrm{gof}_p$ obtained via bootstrap resampling.
}
\label{tab:powerlaws}
\end{table}

\section{Discussion}
\subsection{Limitations}
\subsubsection{Criticality score and spatial propagation limitations}
An important limitation of the criticality score defined in equation \ref{equation:criticalityscore} is that it relies on avalanche statistics computed from the aggregate activity signal, without explicitly accounting for the spatial extent of propagation. As a result, regimes with low vision radius may exhibit apparent critical-like statistics even when the interaction graph is fragmented. In such cases, activity remains confined within small connected components, and the observed scaling reflects locally confined cascades rather than system-wide critical dynamics. Because the interaction graph is a spatial random geometric graph induced by $v_r$, we do not explicitly test small-world shortcuts; future work should compare spatial, small-world, modular, and scale-free graphs to assess how long-range links affect the micro–macro displacement.

This limitation is consistent with the observed deviations from mean-field scaling reported in table~\ref{tab:powerlaws}, where fitted exponents differ from theoretical predictions. Both effects arise from constrained, spatially structured propagation of activity, which violates the homogeneous mixing assumptions underlying classical critical branching processes.

\subsubsection{Finite-size and sampling limitations}
A further limitation of our study concerns the finite size of the simulations and the amount of data available for statistical estimation. Each configuration $(p^{\mathrm{micro}},v_r)$ is evaluated using reservoirs of $64$ neurons per agent, $256$ agents, trajectories of $1000$ timesteps, and $200$ independent runs. While this is sufficient to reveal qualitative trends and identify the emergence of distinct dynamical regimes, it remains limited for accurately characterizing critical behavior, which is inherently sensitive to finite-size effects and sampling variability. In particular, estimates of power-law exponents and goodness-of-fit metrics may be affected by noise, leading to variability across parameter regions. Larger systems, longer simulations, and increased sampling would improve the reliability of the inferred statistics, reduce fluctuations, and allow for a more precise identification of the boundaries of the near-critical regime.

\subsection{Relation to prior work}
\subsubsection{Extended critical regimes}
\begin{figure}[t]
    \centering
    \includegraphics[width=0.95\linewidth]{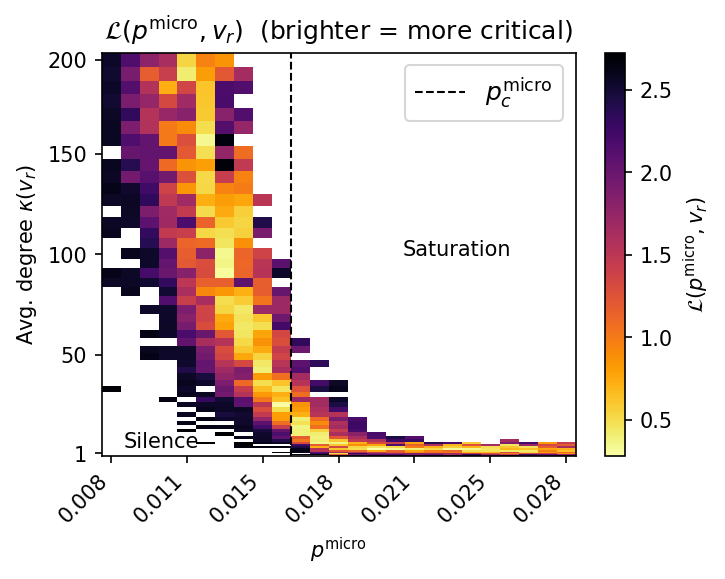}
    \includegraphics[width=0.85\linewidth]{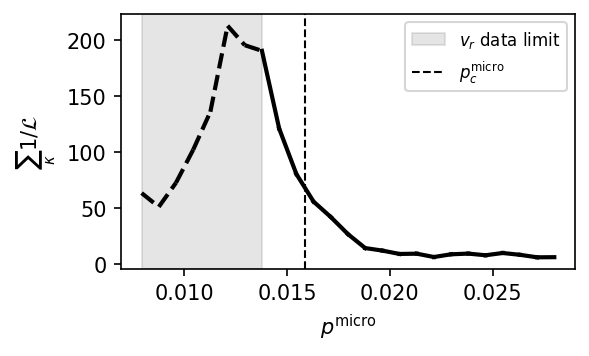}
   \caption{ Top: Criticality score $\mathcal{L}(p^{\mathrm{micro}}, \kappa)$ as a function of $p^{\mathrm{micro}}$ and the average degree $\kappa(v_r)$. Bottom: Integrated criticality $\sum_{\kappa} 1/\mathcal{L}$ as a function of $p^{\mathrm{micro}}$, obtained by summing the inverse criticality score over all active values of $\kappa(v_r)$. Higher values indicate robustness of critical behavior across a wide range of connectivity. The shaded region denotes truncation due to the maximum explored $v_r$ (dashed segments), and the vertical dashed line indicates $p^{\mathrm{micro}}_c$.}
\label{fig:integral}
\end{figure}

To move beyond the specifics of spatial embedding, we re-express the results in terms of the average degree $\kappa$, which captures the effective interaction scale independently of the particular implementation. As shown in Fig.~\ref{fig:integral} (top), plotting $\mathcal{L}(p^{\mathrm{micro}}, \kappa)$ reveals the same qualitative structure observed with $v_r$, indicating that the emergence of critical-like dynamics is governed by network connectivity. This shows that collective near-critical behavior in this model is not tied to a single interaction scale, but can be sustained across a range of connectivity values.

Importantly, the width of this near-critical region depends on $p^{\mathrm{micro}}$. As shown in Fig.~\ref{fig:integral} (bottom), values of $p^{\mathrm{micro}}$ slightly below the nominal micro-level critical point $p_c^{\mathrm{micro}}$ support near-critical behavior across a broader range of $\kappa$, whereas values closer to or above $p_c^{\mathrm{micro}}$ lead to a narrower region or rapid saturation.

This suggests that, in this model, operating slightly below the nominal micro-level critical point yields more robust near-critical behavior under variations in the macroscopic interaction structure. Similar deviations from the nominal critical point have been reported in systems where the effective operating regime depends on external constraints, network structure, or environmental conditions \citep{cramer2020,khajehabdollahi2022,romanczuk_phase_2022}. Importantly, this relation may provide a candidate explanation for why some biological systems embedded in larger interaction networks exhibit subcritical dynamics, such as gene regulatory networks \citep{ramo2006}.

\subsubsection{Cross-scale conditions for collective near-criticality}

Our results show that collective near-critical behavior is not simply inherited from tuning individual reservoirs near their internal critical regime. Although isolated reservoirs can approach their branching critical point, this is not sufficient to generate critical-like collective dynamics. Instead, scale-free-like avalanches emerge only when activity can propagate through the interaction network.

In this system, propagation is mainly controlled by the vision radius $v_r$, which determines effective connectivity between agents. For small $v_r$, the interaction graph is fragmented and activity remains local. As $v_r$ increases, connectivity enables global propagation and system-wide avalanches. For large $v_r$, however, excessive coupling produces sustained activity, moving the system away from the near-critical regime. Thus, macro-level critical-like behavior depends not only on micro-level reservoir dynamics, but on how these dynamics interact with inter-agent coupling.

Preliminary experiments suggest that sensor count has a weaker effect than connectivity on the location of the near-critical regime, though this remains to be tested systematically. One explanation is that sensory inputs are aggregated before perturbing the reservoirs, compressing multiple signals into a bounded drive. Increasing sensor number may therefore not proportionally increase effective excitation, whereas topology directly constrains activity propagation. Overall, these preliminary observations suggest that connectivity and propagation structure play a stronger role than input dimensionality in shaping collective critical-like behavior in this model.

\subsubsection{Tuning collective near-criticality}
Our results show that the macro-level near-critical regime does not necessarily coincide with the intrinsic micro-level critical point. Instead, the location of the collective regime is jointly determined by the internal reservoir dynamics and the structure of inter-agent coupling. In this sense, collective near-criticality is a tunable cross-scale property of the system: by varying $p^{\mathrm{micro}}$ and the interaction range, the macro-level dynamics can be displaced toward subcritical, near-critical, or supercritical regimes. More broadly, this interpretation connects to prior work showing that critical-like regimes can be approached through suitable organizational or parametric constraints, rather than being fixed at a single intrinsic operating point \citep{aguilera2018}.

\section{Conclusions}

In this work, we investigated how critical-like dynamics propagate across scales in a multi-agent system. By combining agents equipped with internal reservoirs operating near the critical branching regime with spatially constrained interactions, we analyzed the conditions under which micro-level dynamics give rise to collective near-critical behavior.

Our results show that near-critical dynamics within individual agents are not sufficient to produce macro-level critical-like avalanche statistics. Instead, collective near-critical behavior depends on whether activity can propagate through the interaction network, which is strongly shaped by topological constraints. In particular, these constraints shift the macro-level near-critical regime relative to the intrinsic critical point of the reservoirs, revealing a mismatch between micro- and macro-level criticality.

More broadly, these results suggest that tuning collective behavior requires jointly considering the dynamical regime of individual agents and the interaction structure governing activity propagation. Future work should examine how adaptive or evolving interaction structures modify this relationship, and whether the same principles extend to richer environments and explicit task settings.

\section{Acknowledgements}
EP acknowledges the support of the European Union (ERC, GROW-AI, 101045094). Views and opinions expressed are however those of the authors only and do not necessarily reflect those of the European Union or the European Research Council. Neither the European Union nor the granting authority can be held responsible for them.

\footnotesize
\bibliographystyle{apalike}
\bibliography{references} 

\end{document}